\documentclass[a4paper,12pt]{article}

\newcommand{\be}{\begin{equation}}
\newcommand{\ee}{\end{equation}}
\newcommand{\bea}{\begin{eqnarray}}
\newcommand{\eea}{\end{eqnarray}}
\newcommand{\bean}{\begin{eqnarray*}}
\newcommand{\eean}{\end{eqnarray*}}

\newcommand{\gapproxeq}{\lower
.7ex\hbox{$\;\stackrel{\textstyle >}{\sim}\;$}}
\newcommand{\lapproxeq}{\lower
.7ex\hbox{$\;\stackrel{\textstyle <}{\sim}\;$}}

\newcommand{\bc}{\begin{center}}
\newcommand{\ec}{\end{center}}
\newcommand{\btab}{\begin{tabular}}
\newcommand{\etab}{\end{tabular}}

\def\beq#1{\begin{equation} \label{#1}}
\def\eeq{\end{equation}}
\newcommand{\ber}{\begin{eqnarray}}
\newcommand{\eer}{\end{eqnarray}}
\newcommand{\bers}{\begin{eqnarray*}}
\newcommand{\eers}{\end{eqnarray*}}

\begin{document}
\begin{titlepage}

\begin{tabbing}
right hand corner using tabbing so it looks neat and in \= \kill
\> {ANL-HEP-PR-02-070}  \\
\> {OUTP-02-35P}    \\
% \> {version 8}   \\
% \>{version numbers: fec=even; harry=odd} \\
% \> {20 August 2002}   \\
\end{tabbing}
\baselineskip=18pt \vskip 0.9in
\begin{center}
{\bf \Large Puzzles in Cabibbo-Suppressed Charm Decays}\\
\vspace*{0.3in}
{\large F. E. Close}\footnote{\tt{e-mail: F.Close@physics.ox.ac.uk}} \\
\vspace{.1in}
{\it Department of Theoretical Physics,
University of Oxford, \\
Keble Rd., Oxford, OX1 3NP, United Kingdom}\\
\vspace{0.1in}
{\large H.J.Lipkin}\footnote{\tt{e-mail: harry.lipkin@weizmann.ac.il}} \\
{\it  Department of Particle Physics
  Weizmann Institute of Science, Rehovot 76100, Israel}\\
{\it School of Physics and Astronomy,
Raymond and Beverly Sackler Faculty of Exact Sciences,\\
        Tel Aviv University, Tel Aviv, Israel }\\
{\it   High Energy Physics Division, Argonne National Laboratory,\\ 
Argonne, IL 60439-4815, USA\\ 
\vspace{0.1in}}
\end{center}

\begin{abstract}

We identify two Cabibbo suppressed $D^+$ decay modes with anomalously high branching ratios
which are not simply explained by any model. All standard model diagrams that
can contribute to these decays are related by symmetries to diagrams for other
decays that do not show any such enhancement. If these high branching ratios are
confirmed by more precise experiments, they may require new physics to explain
them. Anomalies in $D_s$ decays and tests for possible violation of G-parity are discussed.

\end{abstract}

\end{titlepage}

\section {Two anomalously enhanced $D^+$ decays to strange meson pairs} 
 
Two Cabibbo suppressed $D^+$ decay modes have anomalously high branching ratios\cite{PDG}:
\beq{w1}
BR[D^+ \rightarrow  K^*(892)^+\bar K^o] = 3.2 \pm 1.5\%
\eeq
\beq{w2}
BR[D^+ \rightarrow  K^*(892)^+\bar K^*(892)^o] = 2.6 \pm 1.1\%
\eeq

These are the same order as their corresponding Cabibbo allowed branching ratios
\beq{a1}
BR[D^+ \rightarrow  \rho^+\bar K^o] = 6.6 \pm 2.5\%
\eeq
\beq{a2}
BR[D^+ \rightarrow  \rho^+\bar K^*(892)^o] = 2.1 \pm 1.3\%
\eeq

In this letter we show that the
high branching ratios for these Cabibbo suppressed $D^+$ decay modes are not simply explained by any model; 
specifically, all standard model diagrams that
can contribute to these decays are related by symmetries to diagrams for other
decays that do not show any such enhancement. If these high branching ratios are
confirmed by more precise experiments, they may require new physics to explain
them.

We first note that the dominant tree diagrams for the corresponding allowed and
suppressed decays differ only in the weak vertices $c \rightarrow W^+ + s
\rightarrow \rho^+ + s$ and $c \rightarrow W^+ + s
\rightarrow K^*(892)^+ + s$: the hadronization of the strange quark
$s$ and spectator $\bar d$ is common  to both decays. These diagrams should show the expected Cabibbo
suppression which is not observed.

The possibility that $W^+ \to K^*(892)^+$ is somehow similar in strength to
$W^+ \to \rho^+$ can be discounted. First, the topologically similar tau 
decays $BR[\tau^+ \to K^*(892)^+ \nu] \sim 1.3\%$ and 
$BR[\tau^+ \to \rho^+ \nu] \sim 25\%$ exhibit the expected suppression
of the former, as do the corresponding $D^o$ decays

\beq{n1}
BR[D^o \rightarrow  K^*(892)^+\bar K^-] = 0.35 \pm 0.08\%
\eeq
\beq{n2}
BR[D^o \rightarrow  \rho^+\bar K^-] = 10.8 \pm 0.9\%.
\eeq

Indeed, any model of $D^+$ where the charm quark decays as $c \to K^{*+} s$ with the $s$ and the spectator
$\bar d$ combining to make $K^{*+} \bar K^o$ and $K^{*+} \bar K^{*o}$,  will also say that for the $D^o$
the charmed quark decays as $c \to K^{*+} s$ and the $s$ and the spectator $\bar u$ combine to make 
$K^{*+} K^-$ and $K^{*+}  K^{*-}$. However, the corresponding charged and neutral decays are 
empirically very different.
This is independent of the particular model used for the charmed quark decay.

Thus we need something to explain why changing the spectator makes a big difference. 

The above remarks focussed on the dominant tree diagrams. More generally
we now note that when we consider all diagrams contributing to the anomalously
enhanced decays (\ref{w1}-\ref{w2}), each diagram is  related by symmetries to
a very similar diagram for one of the following  decay modes which show the
expected Cabibbo suppression

\beq{s1}
BR[D^+ \rightarrow  K^+\bar K^*(892)^o] = 0.42 \pm 0.05\%
\eeq
\beq{s2}
BR[D^o \rightarrow  K^*(892)^+K^-] = 0.35 \pm 0.08\%
\eeq
\beq{s3}
BR[D^o \rightarrow  K^*(892)^-K^+] = 0.18 \pm 0.01\%
\eeq
\beq{s4}
BR[D^o \rightarrow  K^*(892)^o\bar K^o] < 0.08 \%
\eeq
\beq{s5}
BR[D^o \rightarrow \bar K^*(892)^oK^o] < 0.16 \%
\eeq
\beq{s6}
BR[D^o \rightarrow  K^*(892)^o\bar K^*(892)^o] = 0.14 \pm 0.05\%
\eeq
Our conclusion will be that there is no simple diagram that enhances the suppressed modes
(\ref{w1}-\ref{w2}) without also enhancing others that show no experimental
enhancement.

\section {Analysis of contributing diagrams and comparison with related
decays}

The diagrams contributing to the $D^+$ decays (\ref{w1}-\ref{w2}) can be
classified into two types:

     1. Those in which the spectator antiquark $\bar d$ appears in the final
state and is connected topologically on the same quark line as the initial
$\bar d$; e.g. the tree diagram and the penguin. Diagrams of this type are
shown in Figs. 1 and 2.

     2. Those in which the spectator antiquark is annihilated and then
recreated in a weak annihilation vertex. If no gluons are created from the
initial state before the weak vertex, these diagrams go via an intermediate $u
\bar d$ state which is an eigenstate of G-parity if it has the spin and parity
$0^-$ needed to produce the $K^*(892)^+\bar K^o$ final state. Diagrams of this
typr are shown in Figs. 3 and 4.

We now examine these in turn.  

     {\bf Diagrams of type 1 for $D^+$ decay - see Figs. 1 and 2}:
	 
	 \noindent These can go into the $K^*(892)^+K^-$ and
$K^*(892)^+\bar K^*(892)^-$ decays of the $D^o$ by flipping the isospin of the
spectator $\bar d$ antiquark into a $ \bar u$. Since all interactions of the
spectator antiquark in the diagram involve only isoscalar gluons, except for the
case of the electroweak penguin, the two diagrams have the same contribution.
These diagrams cannot produce an enhancement of the $D^+$ decay without giving
a similar enhancement of the related $D^o$ decays.

Note that this relates diagrams for $D^+$ decays which lead to a pure
I=1 state and $D^o$ decays which lead to a well-defined isospin mixture of I=0
and I=1 states. Our essential assumption is that the spectator quark in both
flavor states interacts only with gluons which are isoscalar, and that there
are no isovector gluons. This is clearly good QCD. But many standard isospin
relations used in weak decay analyses may hold even if the strong interactions
included isovector ``gluon" exchanges. If this is the case, an important QCD
constraint on the decay amplitudes is missing  in their analysis.

Another way to view this possible missing constraint in conventional analyses 
is to write the decays $D \rightarrow K^{**} \bar K$ as
\beq{new}
c + \bar q \rightarrow (s \bar s + u) + \bar q
\eeq
and look at the exchange in the t-channel between the $\bar q$ and the rest
of the system. While isospin invariance allows both isoscalar and
isovector t-channel exchanges, QCD says that only isoscalar t-channel
exchange is allowed.  This of course relates by crossing the I=0 and
I=1 states in the s channel.

     There are two caveats here. One is the electroweak penguin.  Since the
photon couples more strongly to a spectator $\bar u$ than to  a spectator $\bar
d$ it is difficult to see how such a diagram can enhance the $D^+$ decay
relative to the $D^o$ decay.
Furthermore, replacing $\bar d$ by $\bar s$ would give as the nearest analogue for
this electroweak penguin the decays $D_s \to K^{*+} \phi$ and $D_s \to K^{*+} \eta_{s}$.
There are no observations of modes containing
either one or three $K$ that could be consistent with either of these,
which suggests that they are not 
enhanced.
The other caveat is the presence of additional
diagrams in the neutral $D^o$ decays which have no counterpart in the charged
decays; we must consider the possibility that the tree diagrams for $D^o$ are
enhanced but that these extra diagrams might reduce or cancel the contribution  of the enhanced
diagrams. This possibility
is discussed later.

     {\bf Diagrams of type 2 for $D^+$ decays - see Figs. 3 and 4}: 
	 
	 \noindent If no gluons are emitted from the initial state, these
produce a $u \bar d$ state (such as the $\pi(1800)$\cite{lipkinclose}) which decays conserving G-parity.	 
As the $K^*(892)^+\bar K^o$
and $K^+\bar K^*(892)^o$ states are G-conjugates of one another, the G-parity eigenstates are
$\frac{1}{\sqrt{2}}[K^*(892)^+\bar K^o \pm K^+\bar K^*(892)^o ]$ and hence a state of a given G-parity 
will contribute equally to the $K^*(892)^+\bar K^o$
and $K^+\bar K^*(892)^o$ states. However, the observed branching ratios 
(eqs. \ref{w1} and \ref{s1}) differ by almost
an order of magnitude. Although the presence of diagrams with initial gluons
can produce states of opposite G-parity, the two types of diagrams must be fine
tuned so that they nearly cancel for the $K^+\bar K^*(892)^o$ states and
strongly enhance the $K^*(892)^+\bar K^o$. This seems highly 
unlikely.
This possibility could be eliminated by precision data on $D_s$ decays 
where the Cabibbo dominant modes

\beq{f1}
BR[D_s \to K^{*+} \bar K^o] = 4.3 \pm 1.4 \%;BR[D_s \to K^{+} \bar K^{*o}] = 3.3 \pm 0.9 \%;
\eeq
are consistent with being the same. There is however the possiblility of producing these modes by the
color-suppressed topology and a final assessment requires careful determination of the relative importance of these
decay mechanisms. There are, however problems with $D_s$ decays that defy
conventional explanations and may also indicate the presence of new physics 
contributions\cite{phawaii}. We discuss these later. 

     So there seems to be no simple way to explain the large enhancement
of these $D^+$ decays without also implying enhancements for
other modes that do not exhibit such effects empirically. If the experimental
enhancement holds up this may be a key to new physics.

\section {Effects of $D^o$ final state interactions}

\noindent We now examine possible effects of flavor-changing final state interactions
that exist for the $D^o$ decays but not for the $D^+$, in order to eliminate the
possibility that tree diagrams are enhanced for $D^o$ but are being suppressed by destructive
interference with these extra diagrams.

Flipping the isospin of the spectator antiquark in a diagram of type 1
for  $D^+$ decay produces a diagram for  $D^o$ decay containing a $u \bar u$
pair. The $u \bar u$ can be annihilated and changed into a $d \bar d$ or $s \bar s$
by a final state interaction, which has no
counterpart in the diagrams for $D^+$ decay.
To take this into account
we first note that the diagram for $D^o$ decay obtained by flipping the isospin of the
spectator quark is a mixture of isopins $I=1$ and $I=0$. 

Thus, while $D^+ \to \pi^+(1800)$\cite{lipkinclose}, the $D^o$ could couple via $\pi^0(1800)$ and
$\eta(1760)$ say\cite{PDG}. It is then a priori possible that the combination led to a reduction of the charged
kaons and an enhancement of their neutral counterparts (or vice versa). 
The following isospin sum rule relates the amplitudes for the
decay diagrams of type 1, denoted by $A_s$,  to the physical final states and
those to the isospin eigenstates:
\bea
|A_s[D^o \rightarrow K^*(892)^+K^-]|^2 +
|A_s[D^o \rightarrow K^*(892)^o\bar K^o]|^2 = \nonumber \\
=
|A_s[D^o \rightarrow \{K^*(892)\bar K\}_{I=0}]|^2 +
|A_s[D^o \rightarrow \{K^*(892)\bar K\}_{I=1}]|^2
\eea
and analogously for the $K^*\bar K^*$ modes.

We now  note
that the final state in the $D^+$ decay is a pure isospin eigenstate with
$I=1$.  The final state
interactions for the $I=1$ states in the two decays must be the same since the
strong final state interactions are isospin invariant. Thus
  isospin relates the $I=1$ amplitudes for
$D^+$  and $D^o$ decays
\beq
|A_s[D^o \rightarrow \{K^*(892)\bar K\}_{I=1}]|^2 
=(1/2) \cdot
|A_s[D^+ \rightarrow  K^*(892)^+\bar K^o]|^2
\eeq
and so
\bea
|A_s[D^o \rightarrow K^*(892)^+K^-]|^2 +
|A_s[D^o \rightarrow  K^*(892)^o\bar K^o]|^2\kappa
\nonumber \\
=
|A_s[D^o \rightarrow \{K^*(892)\bar K\}_{I=0}]|^2 +
(1/2) \cdot
|A_s[D^+ \rightarrow  K^*(892)^+\bar K^o]|^2
\nonumber \\
\geq
(1/2) \cdot
|A_s[D^+ \rightarrow  K^*(892)^+\bar K^o]|^2
\eea

Upon allowing for the different life times, this can be rewritten
\bea
BR[D^o \rightarrow K^*(892)^+K^-] +
BR[D^o \rightarrow  K^*(892)^o\bar K^o]
\nonumber \\
\geq
(1/2) \cdot
BR[D^+ \rightarrow  K^*(892)^+\bar K^o] \frac{\tau (D^o)}{\tau (D^+)}
\eea
The left hand side is $< 0.6 \%$ 
while the right hand side is
$0.36 \pm 0.39\%$..

A similar analysis can be applied to the $K^*\bar K^*$ decays. Here there is no
datum for $BR[D^o \to   K^*(892)^+\bar K^*(892)^-]$ and so one cannot
definitively rule out such a conspiracy. To satisfy the inequality would
require  $BR[D^o \to   K^*(892)^+\bar K^*(892)^-] \geq  0.38 \pm 0.23\%$. The
neutral modes have\cite{PDG} $BR[D^o \to   K^*(892)^o\bar K^*(892)^o] = 0.14
\pm 0.05\%$,  Thus an improvement in data is needed to definitively rule out
the fine tuning conspiracy for the $\bar{K}K^*$ decays.

 We do not consider here additional diagrams arising from the singly-suppressed
$c$-quark decay $c \rightarrow du\bar d$ which can contribute to $D^o$ 
and not to $D^+$ decays via final state interactions creating an $s\bar s$
pair. A highly unreasonable conspiracy with fine tuning  would be needed 
to produce a cancellation of the anomalously large diagram of type 1. 

It is therefore of interest to check the branching ratios for the transitions
(\ref{w1}-\ref{w2}) and reduce the errors. Using the present data we find:
\beq{wsum}
BR[D^+ \rightarrow  K^*(892)^+\bar K^o] +
BR[D^+ \rightarrow  K^*(892)^+\bar K^*(892)^o] = 5.8  \pm 1.9\%
\eeq
This is still large at three standard deviations.

Established physics seems only able to explain these data by appeal to fine
tuning, which may already be threatened by other data. 

The present data on these anomalous rates come from single experiments
\cite{frabetti,albrecht}. If subsequent experiments show  these large branching
ratios to be in error, then one may need to reconsider the other branching
ratios extracted from the same analyses, (in particular the $D^o \to K^+K^-$
which also appears to be rather larger than expected\cite{PDG}: compared with
the $D^o \to \pi^+ \pi^-$). Conversely, if the large branching ratios are
confirmed with smaller errors, there may be good reason to look for a new
physics explanation. Thus we urge high statistics study of these decays in
dedicated charm production  experiments, such as may be feasible at CLEO-c,
Fermilab or GSI. 

\section {$D_s$ decay puzzles, annihilation and G-parity tests}

We now recall some unresolved puzzles in $D_s$ decays\cite{phawaii}.
These may require new physics contributions related to those required by the
anomalously enhanced singly forbidden decays discusseed above.

\noindent The $D_s$ decay modes  $D_s \rightarrow$ VP and VV show no significant suppression of ``color-suppressed"
$KK^*$ and $K^*K^*$ relative to ``color-favoured" $\phi \pi$ and
$\phi \rho$. Contrast this with $D^o$ decays: the $D^o$ and $D_s$
differ only by spectator quark flavor, but definite color suppression is seen 
in VP and VV $D^o$ decays and not in $D_s$. How can changing the 
flavor of a spectator quark drastically change the degree of color suppression 
in tree diagrams where the spectator quark does not play an active role?

The simplest explanation would be that these $D_s$ decays are driven by
annihilation. If this is the case, then it adds weight to our argument 
at eq. (\ref{f1}). Establishing the pattern and strengths of the annihilation
modes is another critical piece in solving these enigmas.

\noindent The observation of the purely leptonic annihilation decay
$ D_s \rightarrow W^+ \rightarrow \mu^+ \nu_\mu       $
implies the existence of the hadronic annihilation without gluons
$ D_s \rightarrow W^+ \rightarrow u \bar d  \rightarrow (2n+1) \pi$
where the G parity of a J=0 $u \bar d  $ state without additional gluons
forbids the decay into an even number of pions.

We have assumed that G-parity is a good quantum number in our analysis.
It is in principle possible that this is a weak link.
It is therefore of interest to look for:
 
a. The forbidden $D_s \rightarrow 2n \pi$ decays. 
Even upper limits are of interest. Definite evidence would
ndicate some contribution other than the simple annihilation. Note that
this goes beyond the search for the forbidden $\omega \pi$ mode. Any state which
ends up as an even number of pions is forbidden and its observation gives
information about the existence of other annihilation-type diagrams including
gluons or final-state rescattering.
 
b. The allowed $D_s$ decays into states containing an odd number of pions.
These decays must be there somewhere to be consistent with the observed
leptonic decay.
 
c. Decays into states with several neutral pions may be difficult to
detect. States with a single neutral pion can come from allowed odd-G
decays into an $\eta$ and an even number of charged pions. Thus it might be
useful to examine all multipion decays with no more than one neutral and
classify them as follows:
 
(i)    All $D_s$ decays into an odd number of charged pions and nothing else.
 
(ii)    All $D_s$ decays into an odd number of charged pions and an $\eta$.

(iii)   All $D_s$ decays where no $\eta$ is present into an odd number of
charged pions and a single $\pi^o$.
 
The relative numbers of these three inclusive final states might give
information on the validity of the G-parity selection rule that we have assumed
in our analysis.

There are hints of anomalies already, especially in
the $D_s \rightarrow 
\eta \rho$ and $\eta' \rho$ modes.
 
The $\eta'\rho/\eta\rho$ ratio $=0.9 \pm 0.3$ appears to be anomalously large if it is
due to a spectator tree  diagram, where  the p-wave phase space should favour
the $\eta$ significantly and the amplitude ratio is is of order unity with a
value depending in the mixing angle,  For example, we note that
\bea
BR(D_s \rightarrow \eta \l^+ \nu) = 3.5\pm 0.07\% \nonumber \\
BR(D_s \rightarrow \eta' \l^+ \nu) = 0.88\pm 0.03\%.
\eea
However, there is no
clear indication of the  nature of the additional contribution needed. 
Standard model physics implies that 
different parities and G-parities are not mixed by final state interactions.
Positive G-parity is exotic for both parities and cannot have contributions
that go via an intermediate state of a single quark-antiquark pair. The
$\rho\eta$ and $\rho\eta'$ channels are exotic while $\pi\eta$ and $\pi\eta'$
are not.  Yet all states seem to have
anomalously large branching ratios and favor the $\eta'$. There seems to be a
common mechanism independent of the  quantum number of the final state. The
required additional contribution cannot be a simple annihilation  without
additional gluons emitted before annihilation since this produces a G-parity
eigenstate which is right for $\eta' \pi$, but wrong for $\eta' \rho$.

Annihilation with at least two gluons emitted from the initial state and
interaction between these gluons and the $u \bar d$ state produced by an
annihilation diagram could give a small amplitude which might interfere
constructively with the $\eta'$ amplitudes and destructively with $\eta$.
This diagram must also show up in other G-forbidden even-$\pi$ amplitudes. 
Stringent upper limits on this diagram would exclude this mechanism.
 
Annihilation with two gluons emitted from the initial state which then
turn into an $\eta'$ via a hairpin diagram will produce the $\eta'$ rather
than the $\eta$.  
This mechanism can be
compared with $J/\psi \rightarrow \eta' \gamma$ which is
also dominated by a two-gluon hairpin diagram. However, one would also
expect to see this diagram in the semileptonic decay $D_s \rightarrow
\eta' \mu^+ \nu_\mu$, in contradiction to data
 where the $\eta'/\eta$ ratio does not seem to be enhanced.
 
In summary, we advocate a systematic study of Cabibbo suppressed $D$ decays and
of specific $D_s$ channels to test the validity of G-parity and other generally accepted
selection rules in the heavy flavor sector.

%\newpage
\vskip 0.4in
It is a pleasure to thank J. Cumalat, D. Hitlin, V. Sharma and D. Williams  
for helpful discussions and
comments. FEC acknowledges support from EU-TMR program ``EURODAFNE",
contract CT98-0169. 
HJL acknowledges support from the U.S. Department
of Energy, Division of High Energy Physics, 
Contract W-31-109-ENG-38, the United States-Israel
Binational Science Foundation (BSF), Jerusalem, Israel and
the Basic Research Foundation administered by the Israel Academy of 
Sciences and Humanities.

\newpage
 
%Subject: THE TEX file of revised #8732
% Harry -- here are my suggestions to improve the readability

% stuff for the figures
\font\fiverm=cmr5
\input prepictex
\input pictex
\input postpictex
\newdimen\tdim
\tdim=\unitlength
\def\stpltsmbl{\setplotsymbol ({\small .})}
\def\tarrow{\arrow <5\tdim> [.3,.6]}
\newbox\phru
\setbox\phru=\hbox{\beginpicture
\setcoordinatesystem units <\tdim,\tdim>
\stpltsmbl
\setquadratic
\plot
0 0
2.5 3
5 0
7.5 -3
10 0
/
\endpicture}
\def\photonru #1 #2 *#3 /{\multiput {\copy\phru}  at
#1 #2 *#3 10 0 /}
 
\newbox\sru
\setbox\sru=\hbox{\beginpicture
\setcoordinatesystem units <\tdim,\tdim>
\stpltsmbl
\setquadratic
\plot
  0.0   0.0
  4.8   1.5
  7.5   5.0
  7.3   8.5
  5.0  10.0
  2.7   8.5
  2.5   5.0
  5.2   1.5
 10.0   0.0
/
\endpicture}
\def\springru #1 #2 *#3 /{\multiput {\copy\sru}  at
#1 #2 *#3 10 0 /}

{\begin{figure}[htb]
$$\beginpicture
\setcoordinatesystem units <\tdim,\tdim>
\stpltsmbl
\putrule from -25 -10 to 50 -10
\putrule from -25 -10 to -25 30
\putrule from -25 30 to 50 30
\putrule from 50 -10 to 50 30
\plot -25 -20 50 -20 120 -40 /
\plot -25 20 -50 20 /
\plot 50 20 120 40 /
\plot 120 20 50 0 120 -20 /
\put {$c$} [b] at -50 25
\put {$\overline{d}$} [t] at -50 -25
\put {$u$} [l] at 125 40
\put {$\overline{s}$} [l] at 125 20
\put {$s$} [l] at 125 -20
\put {$\overline{d}$} [l] at 125 -40
\put {$\Biggr\}$ $K^*(892)^+$}  [l] at 135 30
\put {$\Biggr\}$ $\bar K^o$} [l] at 135 -30
\setshadegrid span <1.5\unitlength>
\hshade -10 -25 50 30 -25 50 /
\linethickness=0pt
\putrule from 0 0 to 0 60
\endpicture$$
\caption{\label{fig-4}} \hfill   Cabibbo suppressed inactive spectator diagram
\hfill~ \end{figure}}

\centerline{$BR[D^+ \rightarrow  K^*(892)^+\bar K^o] = 3.2 \pm 1.5\%$} 
{\begin{figure}[htb]
$$\beginpicture
\setcoordinatesystem units <\tdim,\tdim>
\stpltsmbl
\putrule from -25 -10 to 50 -10
\putrule from -25 -10 to -25 30
\putrule from -25 30 to 50 30
\putrule from 50 -10 to 50 30
\plot -25 -20 50 -20 120 -40 /
\plot -25 20 -50 20 /
\plot 50 20 120 40 /
\plot 120 20 50 0 120 -20 /
\put {$c$} [b] at -50 25
\put {$\overline{u}$} [t] at -50 -25
\put {$u$} [l] at 125 40
\put {$\overline{s}$} [l] at 125 20
\put {$s$} [l] at 125 -20
\put {$\overline{u}$} [l] at 125 -40
\put {$\Biggr\}$ $K^*(892)^+$}  [l] at 135 30
\put {$\Biggr\}$ $ K^-$} [l] at 135 -30
\setshadegrid span <1.5\unitlength>
\hshade -10 -25 50 30 -25 50 /
\linethickness=0pt
\putrule from 0 0 to 0 60
\endpicture$$
\caption{\label{fig-4}} \hfill   Cabibbo suppressed inactive spectator diagram
\hfill~ \end{figure}}

\centerline{$BR[D^o \rightarrow  K^*(892)^+K^-] = 0.35 \pm 0.08\%$}
 
\centerline{The two diagrams differ only by isospin flip of spectator quark}. 

\pagebreak

{\begin{figure}[htb]
$$\beginpicture
\setcoordinatesystem units <\tdim,\tdim>
\stpltsmbl
\putrule from -25 -30 to 50 -30
\putrule from -25 -30 to -25 30
\putrule from -25 30 to 50 30
\putrule from 50 -30 to 50 30
\plot -25 -20 -50 -20 /
\plot -25 20 -50 20 /
\plot 50 20 120 40 /
\plot 50 -20 120 -40 /
\springru 50 0 *3 /
\plot 120 20 90 0 120 -20 /
\put {$c$} [b] at -50 25
\put {$\overline{d}$} [t] at -50 -25
\put {$u$} [l] at 125 40
\put {$\overline{s}$} [l] at 125 20
\put {$s$} [l] at 125 -20
\put {$\overline{d}$} [l] at 125 -40
\put {$\Biggr\}$ $K^*(892)^+$} [l] at 135 30
\put {$\Biggr\}$ $\bar K^o$} [l] at 135 -30
\put {$G$} [t] at 70 -5
\setshadegrid span <1.5\unitlength>
\hshade -30 -25 50 30 -25 50 /
\linethickness=0pt
\putrule from 0 0 to 0 60
\endpicture$$
\caption{\label{fig-2}} \hfill Annihilation Diagram. $G$ denotes any number of
gluons. \hfill~ \end{figure}}

\centerline{$BR[D^+ \rightarrow  K^*(892)^+\bar K^o] = 3.2 \pm 1.5\%$} 

{\begin{figure}[htb]
$$\beginpicture
\setcoordinatesystem units <\tdim,\tdim>
\stpltsmbl
\putrule from -25 -30 to 50 -30
\putrule from -25 -30 to -25 30
\putrule from -25 30 to 50 30
\putrule from 50 -30 to 50 30
\plot -25 -20 -50 -20 /
\plot -25 20 -50 20 /
\plot 50 20 120 40 /
\plot 50 -20 120 -40 /
\springru 50 0 *3 /
\plot 120 20 90 0 120 -20 /
\put {$c$} [b] at -50 25
\put {$\overline{d}$} [t] at -50 -25
\put {$u$} [l] at 125 40
\put {$\overline{s}$} [l] at 125 20
\put {$s$} [l] at 125 -20
\put {$\overline{d}$} [l] at 125 -40
\put {$\Biggr\}$ $K^+$} [l] at 135 30
\put {$\Biggr\}$ $\bar K^*(892)^o$} [l] at 135 -30
\put {$G$} [t] at 70 -5
\setshadegrid span <1.5\unitlength>
\hshade -30 -25 50 30 -25 50 /
\linethickness=0pt
\putrule from 0 0 to 0 60
\endpicture$$
\caption{\label{fig-2}} \hfill Annihilation Diagram. $G$ denotes any number of
gluons. \hfill~ \end{figure}}

\centerline{$BR[D^+ \rightarrow  K^+\bar K^*(892)^o] = 0.42 \pm 0.05\%$}

\centerline{Two final states are $G$-conjugate}.

\centerline{Any G-conserving strong interaction must fine-tune}

\centerline{G-even and G-odd amplitudes to interfere}

\centerline{constructively for $D^+ \rightarrow  K^*(892)^+\bar K^o$}

\centerline{ destructively for $D^+ \rightarrow  K^+\bar K^*(892)^o$}.
\pagebreak

 \end{document}